# Energy dispersion relation for negative refraction (NR) materials

Y. Ben-Aryeh

Physics Department, Technion Israel of Technology, Haifa 32000, Israel

E-mail address: phr65yb@physics.technion,ac.il;  Fax:972 4 8295755

**ABSTRACT**

A general energy dispersion relation is developed for metamaterials having the negative-refraction (NR) property. It is shown that absorption effects are involved with NR phenomena, and the conditions under which NR occurs are discussed. Simple equations for NR are developed by using Lorentzian models..

**1. Introduction**

We describe briefly the idea presented by Veselago [1] and then show the problem related to energy dispersion relations [2].

Assuming a plane monochromatic wave in which all quantities are proportional to $\exp[i(kz - \omega t)]$ we get from Maxwell equations

$$\vec{k} \times \vec{E} = \frac{\omega}{c}\mu\vec{H} \quad ; \quad \vec{k} \times \vec{H} = -\frac{\omega}{c}\varepsilon\vec{E} \quad . \tag{1}$$

Here $\vec{E}$ and $\vec{H}$ are the electric and magnetic fields, respectively, $\varepsilon$ and $\mu$ are the electric permittivity and magnetic permeability, respectvily, $\omega$ is the frequency and $\vec{k}$ is the wave vector. It can be seen from these equations that if $\varepsilon < 0$ and $\mu < 0$ then $\vec{E}, \vec{H}$ and $\vec{k}$ form a left handed system different from the right handed system for the common case of $\varepsilon > 0$ and $\mu > 0$. Assuming that $\hat{z}$ is the unit vector along the Poynting vector, representing the flux of energy, one finds that for materials with negative values of $\mu$ and $\varepsilon$ the wavevector $\vec{k} = \frac{n\omega}{c}\hat{z}$ appears with negative refraction index $n$ [3,4]. As follows from Snell's law the refracted beam from ordinary material ($n \geq 1$) into Negative-refraction (NR) material ($n < 0$) will emerge on the other side of the normal relative to ordinary material. While negative permittivity is obtained quite easily in plasma [5] it is quite difficult to obtain negative magnetic permeability. Since negative index of refraction shows interesting optical phenomena [6-8] large effort has been spent in obtaining a negative magnetic permeability in addition to a negative permittivity in special materials known as metamaterials [9-10] with the NR property. NR has been realized experimentally in various experiments [11-13].

The treatment of NR by analyzing phase velocities does not show the complete nature of these phenomena. Negative values for $\varepsilon$ and/or $\mu$ can appear in metamaterials only if a corresponding energy dispersion relation is valid. The dispersion for the EM (electromagnetic) energy $W$ has been described as [1,2]:

$$W = \frac{1}{2}\frac{\partial(\varepsilon\omega)}{\partial\omega}E^2 + \frac{1}{2}\frac{\partial(\mu\omega)}{\partial\omega}H^2 \quad , \tag{2}$$

where the derivatives in Eq. (2) are relative to frequency. The idea following from the use of Eq. (2) for metamaterials is that we can use in this equation real values for $\varepsilon$ and $\mu$ with negative values $\varepsilon_\Re \equiv \operatorname{Re}\varepsilon < 0$ and $\mu_\Re \equiv \operatorname{Re}\mu < 0$, only when $\frac{\partial(\varepsilon_\Re)}{\partial\omega}$ and $\frac{\partial(\mu_\Re)}{\partial\omega}$ are positive so that the total energy $W$ becomes positive. Eq. (2) has been derived by Landau and Lifshitz [2] under the assumption that $\varepsilon(\omega)$ and $\mu(\omega)$ are real. Usually NR is obtained near resonances where the real parts of the permittivity and permeability change rapidly as function of frequency. As is well known, the real and imaginary parts of the permittivity and permeability are related in such regions by Kramers-Kronig (KK) relations and such relations are manifested by Hilbert transforms (HT) [14-18]. Therefore, the relation between the imaginary and real parts of these functions cannot be ignored and the assumption of real values for $\varepsilon$ and $\mu$ is not valid for metamaterials. Also we will show that near such resonances there are spectral regions where $\varepsilon_\Re$ and $\mu_\Re$ are negative including negative values for their derivatives. Therefore, in such spectral regions Eq. (2) canot lead to positive energy. The possibility to use in Eq. (2) complex values $\varepsilon_\Re + i\varepsilon_\Im$ and $\mu_\Re + i\mu_\Im$, where in a short notation the subscripts $\Re$ and $\Im$ denote real and imaginary parts, respectively, does not solve the problem as $W$ should be a real quantity. Theses problems indicate that the use of Eq.(2) should be generalized.

For developing the general energy dispersion relation we describe in Section 2 the dependence of $\varepsilon$ and $\mu$ on frequency $\omega$ and show the relations of such functions to HT [14-18]. Then, we derive in Section 3 the energy dispersion relation for metamaterials which will generalize the use of Eq. (2). We will show that there is a certain threshold for which Eq. (2) will give positive energy. In Section 4 we analyze the implications of the energy dispersion relations to Lorentzian models for the electric [5] and magnetic [19-24] susceptibilities. In Section 5 we summarize our results and conclusions.

## 2. The dependence of $\varepsilon$ and $\mu$ on frequency related to Hilbert transforms (HT)

Kramers-Kronig (KK) relations and their mathematical manifestation by HT can be applied to dielectric and magnetic materials that absorb light. The dielectric optical properties of such materials are represented by the equations for a complex displacement electric field $\vec{D}$ and polarization $\vec{P}$ [25,26]:

$$\vec{P} = \varepsilon_0 \chi \vec{E} \quad ; \quad \vec{D} = \varepsilon \vec{E} = \varepsilon_0 \varepsilon_r \vec{E} \quad ; \quad \varepsilon_r = (1+\chi) \quad ; \quad \chi = \chi' + i\chi'' \quad , \tag{3}$$

corresponding to a complex permittivity $\varepsilon$ and a complex electric susceptibility $\chi$. The magnetic optical properties are represented by the equations for a complex magnetic field $\vec{B}$ [26] and magnetic polarization $\vec{M}$:

$$\vec{M} = \mu_0 \xi \vec{H} \quad ; \quad \vec{B} = \mu \vec{H} = \mu_0 \mu_r \vec{H} \quad ; \quad \mu_r = (1+\xi) \quad ; \quad \xi = \xi' + i\xi'' \tag{4}$$

corresponding to a complex permeability $\mu$ and a complex magnetic susceptibility $\xi$. We use here the isotropic assumption but for more general cases $\varepsilon$ and $\chi$ will be tensors [27]. The KK relations for the dielectric functions using a linear model are given as [15]:

$$\begin{aligned} \chi_\Re(\omega) &= \text{Re}(\varepsilon_r(\omega)) - 1 = \frac{2}{\pi} \int_0^\infty \frac{\text{Im}(\varepsilon_r(\omega'))\omega' d\omega'}{\omega'^2 - \omega^2} \quad , \\ \chi_\Im(\omega) &= \text{Im}(\varepsilon_r(\omega)) = -\frac{2\omega}{\pi} \int_0^\infty \frac{[\text{Re}(\varepsilon_r(\omega')) - 1] d\omega'}{\omega'^2 - \omega^2} \end{aligned} \tag{5}$$

where $\chi_\Re(\omega)$ and $\chi_\Im(\omega)$ are the real and imaginary parts of the electric susceptibility $\chi$. In a similar way one gets [15]:

$$\begin{aligned} \xi_\Re(\omega) &= \text{Re}(\mu_r(\omega)) - 1 = \frac{2}{\pi} \int_0^\infty \frac{\text{Im}(\mu_r(\omega'))\omega' d\omega'}{\omega'^2 - \omega^2} \quad ; \\ \xi_\Im(\omega) &= \text{Im}(\mu_r(\omega)) = -\frac{2\omega}{\pi} \int_0^\infty \frac{\text{Re}(\mu_r(\omega') - 1) d\omega'}{\omega'^2 - \omega^2} \end{aligned} \tag{6}$$

where $\xi_\Re(\omega)$ and $\xi_\Im(\omega)$ are the real and imaginary parts of the magnetic susceptibility. One should take into account that $\chi_\Re(\omega)$ and $\xi_\Re(\omega)$ are antisymmetric functions in $\omega$ while $\chi_\Im(\omega)$ and $\xi_\Im(\omega)$ are symmetric so that the HT manifested in Eqs. (5) and (6) relate the symmetric functions with the antisymmetric ones. In Eqs.

(5) and (6) we have omitted the need for using Principal Value (P.V.) [26] as we have assumed that the spectral region of $\omega$ is very far from $\omega = 0$ [15].

A simple Lorentzian model of the electric polarization density $\vec{P}$ and the corresponding permittivity $\varepsilon$ can be related to a damped and driven harmonic oscillator [5]. Assuming a monchromatice time dependence $E(t) = E_0 \exp(-i\omega t)$, and $P(t) = P_0 \exp(-i\omega t)$ one gets in this model:

$$P_0 = \varepsilon_0 \chi(\omega) E_0 \quad ; \quad \chi(\omega) = \chi_0 \frac{\omega_0^2}{\omega_0^2 - \omega^2 - i\omega\gamma} \quad ; \quad \varepsilon = \varepsilon_0 (1 + \chi(\omega)) \tag{7}$$

where $\omega_0$ is the natural frequency of the oscillator. Such model describes certain oscillations in plasma where

$$\chi_0 = \frac{Ne^2}{m\varepsilon_0 \omega_0^2} \quad . \tag{8}$$

Here $e$ and $m$ are the charge and effective mass of the electron, respectively, and $N$ is the number of charges per unit volume.

One should take into account that the above functions are valid only for linear systems. Using linear models for metamaterials a magnetic Lorentzian model has been developed in various works [19-24] leading to the relation:

$$\mu_r(\omega) = 1 - \frac{F\omega_0^2}{\omega^2 - \omega_0^2 + i\omega\Gamma} \quad , \tag{9}$$

where $\omega_0$ is the resonance frequency for the permeability, $\Gamma$ represent losses and $F$ represents the strength of the Lorentzian mode interactions. One should notice that the magnetic susceptibility $\xi(\omega) = \mu_r(\omega) - 1$ has a very similar form to the electric susceptibility $\chi(\omega)$ derived by the electric Lorentzian model in Eq. (7). While the derivation of the electric Lorentzian model for a driven and damped oscillator is very simple, derivations of magnetic Lorentzian models are quite complicated. [see e.g.

21]. In the present study we assume a resonance frequency $\omega_0$ which is usually assumed to be given for metamaterials so that we do not treat here the Drude model [5] for which $\omega_0 = 0$. The resonance frequency in magnetic Lorentzian model is usually related to capacitance and inductive elements while the losses are related to resistance elements, and such relations can be realized in special structures known as 'split ring resonators' [19-26]. The explicit calculations for different metamaterials give different evaluations for the constants $\omega_0$, $\Gamma$ and $F$ but usually lead to magnetic susceptibility of the form (9) or a similar one. As our interest in the present paper is in energy dispersion relations we will study in the next Section the general relations between magnetic and electric susceptibilities and the energy dispersion relations.

## 3. Energy dispersion relation for negative refraction materials

The change of electric and magnetic energies per unit volume and unit time $\frac{\partial U}{\partial t}$ is given by [2]:

$$\frac{\partial U}{\partial t} = \frac{1}{2}\left(\vec{E}\cdot\frac{\partial \vec{D}^*}{\partial t} + \vec{E}^*\cdot\frac{\partial \vec{D}}{\partial t}\right) + \frac{1}{2}\left(\vec{H}\cdot\frac{\partial \vec{B}^*}{\partial t} + \vec{H}^*\cdot\frac{\partial \vec{B}}{\partial t}\right) = \text{Re}\left(\vec{E}^*\cdot\frac{\partial \vec{D}}{\partial t}\right) + \text{Re}\left(\vec{H}^*\cdot\frac{\partial \vec{B}}{\partial t}\right)$$

(10)

Such relation follows from calculations of the Poynting vector where the electric and magnetic fields are complex vectors [2,26]. By assuming a narrow band of frequencies for the electric and magnetic fields they can be given as:

$$\vec{E}(t) = \vec{E}_0(t)\exp(-i\omega t) \quad , \quad \vec{E}_0(t) = \int_\alpha \vec{E}_{0,\alpha}\exp(-i\alpha t)d\alpha \quad ,$$
$$\vec{E}_{0,\alpha} = \frac{1}{2\pi}\int_t E_0(t)\exp(i\alpha t)dt$$

(11)

$$\vec{H}(t) = \vec{H}_0(t)\exp(-i\omega t) \quad ; \quad \vec{H}_0(t) = \int_\alpha \vec{H}_{0,\alpha}\exp(-i\alpha t)d\alpha,$$

$$\vec{H}_{0,\alpha} = \frac{1}{2\pi}\int_t \vec{H}_0(t)\exp(i\alpha t)dt \tag{12}$$

$$\alpha \ll \omega \quad ; \quad \omega' = \omega + \alpha \quad ; \quad \frac{\partial \omega'}{\partial t} = \alpha. \tag{13}$$

Here $\omega$ is the central 'carrier frequency' and the integrals over $\alpha$ describe integrals over a narrow frequency distribution around the central frequency $\omega$. $\vec{E}_{0\alpha}$ and $\vec{H}_{0\alpha}$ are components in frequency space of the Fourier transform of $\vec{E}_0(t)$ and $\vec{H}_0(t)$, respectively.

We develop $\varepsilon(\omega')$ and $\mu(\omega')$ up to first order in $\Delta\omega' = \alpha$. Then we get:

$$\varepsilon(\omega') = \varepsilon(\omega) + \left(\frac{\partial \varepsilon(\omega')}{\partial \omega'}\right)_{\omega'=\omega} \alpha, \tag{14}$$

$$\mu(\omega') = \mu(\omega) + \left(\frac{\partial \mu(\omega')}{\partial \omega'}\right)_{\omega'=\omega} \alpha. \tag{15}$$

Using these approximations the displaced electric field $\vec{D}$ and the induced magnetic field $\vec{B}$ are given as

$$\vec{D} = \int_\alpha \left\{\vec{E}_{0\alpha}\exp(-i(\omega+\alpha)t)\left[\varepsilon(\omega) + \left(\frac{\partial \varepsilon(\omega')}{\partial \omega'}\right)_{\omega'=\omega}\alpha\right]\right\}d\alpha \tag{16}$$

$$\vec{B} = \int_\alpha \left\{\vec{H}_{0\alpha}\exp(-i(\omega+\alpha)t)\left[\mu(\omega) + \left(\frac{\partial \mu(\omega')}{\partial \omega'}\right)_{\omega'=\omega}\alpha\right]\right\}d\alpha \tag{17}$$

Performing the time derivatives $\frac{\partial \vec{D}}{\partial t}$ and $\frac{\partial \vec{B}}{\partial t}$ and neglecting terms of order $\alpha^2$ we get:

$$\frac{\partial \vec{D}}{\partial t} = \int_\alpha \left\{\vec{E}_{0\alpha}\exp(-i(\omega+\alpha)t)\left(-i\omega\left[\varepsilon(\omega) + \left(\frac{\partial \varepsilon(\omega')}{\partial \omega'}\right)_{\omega'=\omega}\alpha\right] - i\alpha\varepsilon(\omega)\right)\right\}d\alpha, \tag{18}$$

$$\frac{\partial \vec{B}}{\partial t} = \int_\alpha \left\{ \vec{H}_{0\alpha} \exp(-i(\omega+\alpha)t) \left( -i\omega \left[ \mu(\omega) + \left( \frac{\partial \varepsilon(\omega')}{\partial \omega'} \right)_{\omega'=\omega} \alpha \right] - i\alpha\mu(\omega) \right) \right\} d\alpha \quad . \quad (19)$$

For the first term on the right side of Eqs. (18) and (19) we can use respectively, the relations:

$$\int_\alpha \left\{ \vec{E}_{0\alpha} \exp(-i(\omega+\alpha)t)(-i\omega\varepsilon(\omega)) \right\} d\alpha = -i\omega\varepsilon(\omega)\vec{E}(t) \quad , \quad (20)$$

$$\int_\alpha \left\{ \vec{H}_{0\alpha} \exp(-i(\omega+\alpha)t)(-i\omega\mu(\omega)) \right\} d\alpha = -i\omega\mu(\omega)\vec{H}(t) \quad . \quad (21)$$

In Eqs. (20) and (21), $\omega$, $\varepsilon(\omega)$ and $\mu(\omega)$ have been taken out of the integral and Eqs. (11-12) have been used, respectively.

For the second and third terms on the right side of Eqs. (18) and (19) we use, respectively, the relations:

$$\left( -i\alpha \left[ \left( \frac{\partial \varepsilon(\omega')}{\partial \omega'} \right)_{\omega'=\omega} \right] - i\alpha\varepsilon(\omega) \right) = -i\alpha \frac{\partial}{\partial \omega'} [\omega'\varepsilon(\omega')]_{\omega'=\omega} \quad , \quad (22)$$

$$\left( -i\alpha \left[ \left( \frac{\partial \mu(\omega')}{\partial \omega'} \right)_{\omega'=\omega} \right] - i\alpha\mu(\omega) \right) = -i\alpha \frac{\partial}{\partial \omega'} [\omega'\mu(\omega')]_{\omega'=\omega} \quad . \quad (23)$$

Inserting Eqs. (20) and (22) into Eq. (18) and using the relation $\frac{\partial}{\partial t}(\vec{E}_0(t)) = -i\alpha(\vec{E}_0(t))$ we get

$$\frac{\partial \vec{D}}{\partial t} = -i\omega\varepsilon(\omega)\vec{E}(t) + \frac{\partial \vec{E}_0}{\partial t} \frac{\partial}{\partial \omega'} [\omega'\varepsilon(\omega')]_{\omega'=\omega} \quad . \quad (24)$$

Inserting Eqs. (21) and (23) into Eq. (19) and using the relation $\frac{\partial}{\partial t}(\vec{H}_0(t)) = -i\alpha(\vec{H}_0(t))$ we get:

$$\frac{\partial \vec{B}}{\partial t} = -i\omega\mu(\omega)\vec{H}(t) + \frac{\partial \vec{H}_0(t)}{\partial t} \frac{\partial}{\partial \omega'} [\omega'\mu(\omega')]_{\omega'=\omega} \quad . \quad (25)$$

Inserting Eqs. (24-25) into Eq. (10) and using Eqs. (11-12) we finally get:

$$\frac{\partial U}{\partial t} = \frac{1}{2}\left[\frac{\partial}{\partial t}|\vec{E}(t)|^2\right]\frac{\partial}{\partial \omega'}[\omega' \varepsilon_{\Re}(\omega')]_{\omega'=\omega} + \frac{1}{2}\left[\frac{\partial}{\partial t}|\vec{H}(t)|^2\right]\frac{\partial}{\partial \omega'}[\omega' \mu_{\mathbb{R}}(\omega')]_{\omega'=\omega}$$
$$+\omega\varepsilon_{\Im}|\vec{E}(t)|^2 + \omega\mu_{\Im}(\omega)|\vec{E}(t)^2| \qquad (26)$$

Here we have used the relations $|\vec{E}_0(t)|^2 = |\vec{E}(t)|^2$, $|\vec{H}_0(t)|^2 = |\vec{H}(t)|^2$. We remind that the subscripts $\Re$ and $\Im$ denote the real and imaginary parts where according to Eqs (3-4) we have, respectively:

$$\varepsilon_{\Re} = \varepsilon_0(1+\chi') \quad ; \quad \varepsilon_{\Im} = \varepsilon_0 \chi'' \quad ; \quad \mu_{\Re} = \mu_0(1+\xi') \quad ; \quad \mu_{\Im} = \mu_0 \xi'' \qquad (27)$$

The first two terms on the right side of Eq. (26) represent the change in the EM energy which enters into the metamaterial while the additional two terms represent the absorbed energy, both per unit time and per unit volume. Following the above approximations, in the derivatives according to $\omega'$ appear only the real parts $\varepsilon_{\Re}$ and $\mu_{\mathbb{R}}$ while in the last two terms of Eq. (26), representing absorption, appear only the imaginary parts $\varepsilon_{\Im}$ and $\mu_{\Im}$.

One should take care about the following points. Assuming zero electric and magnetic fields at a certain initial time and integrating the first two terms on the right side of Eq.(23) till time $t$ the result will be equivalent to that of Eq. (2) under the conditions $\frac{\partial}{\partial \omega}[\omega\varepsilon_{\mathbb{R}}(\omega)] > 0$ and $\frac{\partial}{\partial \omega}[\omega\mu_{\mathbb{R}}(\omega)] > 0$. For spectral regions for which $\frac{\partial}{\partial \omega}[\omega\varepsilon_{\mathbb{R}}(\omega)] < 0$ and $\frac{\partial}{\partial \omega}[\omega\mu_{\mathbb{R}}(\omega)] < 0$, Eq. (2) leads to negative energy $W$ which cannot be true. In such cases we get according to Eq. (26):

$$\left[\frac{\partial}{\partial t}|\vec{E}(t)|^2\right] = \left[\frac{\partial}{\partial t}|\vec{H}(t)|^2\right] = 0 \quad ; for \quad \frac{\partial}{\partial \omega}[\omega\varepsilon_{\mathbb{R}}(\omega)] < 0; \frac{\partial}{\partial \omega}[\omega\mu_{\mathbb{R}}(\omega)] < 0 \qquad (28)$$

We find that under the condition of Eq. (28) all the EM incident on the metamaterial is reflected ,i.e., the metamateial becomes opaque, with vanishing EM fields in this material. There are, however, extensive spectral regions in which the restriction (28) does not apply and NR can be implemented. In such spectral regions one should take into account that a part of the energy supplied to the electric and magnetic fields in the metamaterials is absorbed as given by the additional terms in Eq. (26).

The implications of the more general energy dispersion relations for materials satisfying the Lorentzian spectral profiles will be treated in the next Section

## 3. Energy dispersion relation for metamaterials satisfying Lorentzian spectral profiles

In order to use the energy dispersion relation to metamaterials with Lorentzian spectral profiles we need to separate $\varepsilon$ and $\mu$ into their real and imaginary parts. Using Eq. (7) the real and imaginary parts of the electric susceptibility are given as

$$\chi_\Re(\omega) = \chi_0 \frac{\omega_0^2\left(\omega_0^2 - \omega^2\right)}{\left(\omega_0^2 - \omega^2\right)^2 + \omega^2\gamma^2} \quad ; \quad \chi_\Im(\omega) = \chi_0 \frac{\omega_0^2 \omega\gamma}{\left(\omega_0^2 - \omega^2\right)^2 + \omega^2\gamma^2} \quad . \tag{29}$$

We find that $\chi_\Re(\omega)$ is negative when $\omega > \omega_0$ and $\varepsilon_\Re = \varepsilon_0(1+\chi_\Re)$ becomes negative at a certain spectral range when $\chi_\Re < -1$. In order to estimate the spectral range where $\varepsilon_\Re$ is negative and to estimate the value $\frac{\partial \varepsilon_\Re(\omega)}{\partial \omega}$ in this spectral range we introduce the approximations:

$$\omega \simeq \omega_0 \quad ; \quad \omega_0^2 - \omega^2 = (\omega_0 - \omega)(\omega_0 + \omega) \simeq 2\omega_0(\omega_0 - \omega) \quad . \tag{30}$$

These approximations are valid as long as $|\omega_0 - \omega| \ll \omega_0$. Under these approximations

$$\varepsilon_\Re \simeq \varepsilon_0\left[1 + \chi_0 \frac{(\omega_0 - \omega)(\omega_0/2)}{(\omega_0 - \omega)^2 + (\gamma/2)^2}\right] \quad , \tag{31}$$

$$\frac{\partial \varepsilon_\Re}{\partial \omega} \simeq \varepsilon_0 \left[ -\frac{(\chi_0 \omega_0 / 2)}{(\omega_0 - \omega)^2 + (\gamma/2)^2} + \frac{\chi_0 (\omega_0 - \omega)^2 \omega_0}{\left[(\omega_0 - \omega)^2 + (\gamma/2)^2\right]^2} \right] \quad , \quad (32).$$

$$\varepsilon_\Im \simeq \left[ \varepsilon_0 \chi_0 \frac{(\gamma \omega_0 / 4)}{(\omega_0 - \omega)^2 + (\gamma/2)^2} \right]. \quad (33)$$

One should notice that $\varepsilon_\Im$ is symmetric in $\omega$ while $\varepsilon_\Re - \varepsilon_0$ is antisymmetric. According to Eq. (31) $\varepsilon_\Re$ becomes negative when $\omega > \omega_0$ and $(\omega_0 - \omega)^2 + (\gamma/2)^2$ is small relative to $\chi_0 |\omega_0 - \omega|(\omega_0/2)$. However, $\frac{\partial \varepsilon_\Re}{\partial \omega}$ becomes negative when $|\omega_0 - \omega| < \gamma/2$ and under this conditions the electric field does not penetrate the metamaterial. When $\frac{\partial \varepsilon_\Re}{\partial \omega}$ becomes positive the value of $\frac{\partial \varepsilon_\Re}{\partial \omega} \omega$ increases quite rapidly as function of frequency and NR is implemented.

Performing similar calculations for the magnetic Lorentzian profiles, under the same approximation, we get

$$\mu_\mathbb{R}(\omega) = \mu_0 \left[ 1 + F \frac{(\omega_0 - \omega)(\omega_0/2)}{(\omega_0 - \omega)^2 + (\Gamma/2)^2} \right] \quad , \quad (34)$$

$$\frac{\partial \mu_\Re(\omega)}{\partial \omega} = \mu_0 \left[ -\frac{F\omega_0/2}{(\omega_0 - \omega)^2 + (\Gamma/2)^2} + \frac{F\omega_0(\omega_0 - \omega)^2}{\left[(\omega_0 - \omega)^2 + (\Gamma/2)^2\right]^2} \right] \quad , \quad (35)$$

$$\mu_\Im(\omega) = \mu_0 \left[ F \frac{(\Gamma \omega_0/4)}{(\omega_0 - \omega)^2 + (\Gamma/2)^2} \right]. \quad (36)$$

Here again $\mu_\Re$ becomes negative when $(\omega_0 - \omega)$ is negative and $(\omega_0 - \omega)^2 + (\Gamma/2)^2$ is small relative to $F|\omega_0 - \omega|(\omega_0/2)$. $\frac{\partial \mu_\Re(\omega)}{\partial \omega}$ becomes, however, negative when $|\omega_0 - \omega| < \Gamma/2$ and under this condition the magnetic field does not penetrate the metamaterial. It is also easy to show that beyond these special spectral regions there are extensive regions for which $\varepsilon_\Re(\omega)$ and $\mu_\Re(\omega)$ are negative while $\frac{\partial(\varepsilon_\Re\omega)}{\partial\omega}$ and $\frac{\partial(\mu_\Re\omega)}{\partial\omega}$ become positive and relatively large so that in these regions the incident EM field penetrates into the metamaterial and shows the interesting NR phenomena. Even in such cases a part of the incident energy is absorbed as described by Eqs. (26).

## 5. Summary and Conclusion

The usual energy dispersion relation, given by Eq. (2), has been generalized by analyzing Poynting vector effects for EM signal, with narrow band of frequencies, following the use of Eq. (10). The derivation of Eqs, (26) shows two important effects which should be taken into account in the implementation of NR phenomena: a) Absorption effects per unit time and unit volume are described in Eq. (26). b) It follows from Eq.(26) that under the conditions $\frac{\partial}{\partial\omega'}[\omega'\varepsilon_\Re(\omega')]_{\omega'=\omega} < 0$, $\frac{\partial}{\partial\omega'}[\omega'\mu_\Re(\omega')]_{\omega'=\omega} < 0$ the EM fields cannot penetrate the metamaterials as the only solution for this equation under these conditions is $|\vec{E}(t)|^2 = |\vec{H}(t)|^2 = 0$. Beyond these spectral regions positive energy in the metamaterials can be realized in agreement of Eq. (2) or with the more general energy dispersion relation of Eq.(26).

In order to realize NR phenomena one has to apply the incident EM waves with frequencies near resonances in order to get large values of $\frac{\partial(\varepsilon_\Re)}{\partial\omega}$ and $\frac{\partial(\mu_\Re)}{\partial\omega}$. Such resonances can be described quite often by Lorentzian models. The different parameters needed for NR phenomena have been estimated by developing approximate equations using Lorentzian models.